\documentstyle[ltwol,epsfig]{article}

\def\be{\begin{equation}}
\def\ee{\end{equation}}
\def\bea{\begin{eqnarray}}
\def\eea{\end{eqnarray}}
\def\Dzero{D$\emptyset$}
\def\GeV{{\rm GeV}}
\def\TeV{{\rm TeV}}
\def\xbj{x_{\rm Bj}}
\def\cG{{\cal G}}
\def\qbar{\bar{q}}

\begin{document}
\title{\vspace*{-4\baselineskip}
\hfill {\rm UR-1556}\\
\hfill {\rm ER/40685/925}\\
\hfill {\rm DTP/98/82}\\
\hfill {\rm November 1998}\\
\vspace*{\baselineskip}
A BFKL MONTE CARLO APPROACH TO JET PRODUCTION\\ AT HADRON-HADRON
AND LEPTON-HADRON COLLIDERS}

\author{L.H.~ORR}

\address{Department of Physics and Astronomy, University of
Rochester, Rochester NY 14627-0171, USA\\E-mail: orr@pas.rochester.edu}   

\author{W.J.~STIRLING}

\address{Departments of Physics and Mathematical Sciences, University of
Durham, Durham DH1 3LE, UK\\E-mail: W.J.Stirling@durham.ac.uk}  


\twocolumn[\maketitle\abstracts{The production of a pair of jets with 
large rapidity separation in 
hadron-hadron collisions, and of forward jets in deep inelastic scattering, 
can in principle be used to test the predictions of the BFKL equation. 
However in practice kinematic constraints lead to a strong suppression of 
BFKL effects for these processes. This is illustrated using a BFKL Monte 
Carlo approach.\footnote{Presented by LHO at the 
XXIX International Conference on High Energy Physics, Vancouver, B.C.,
July 23-29, 1998.}  }]

\section{Introduction}

\begin{figure}
\psfig{figure=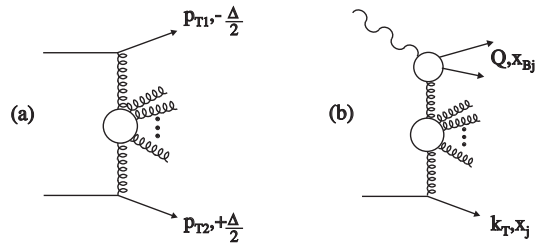,height=3.5in,angle=270}
\vskip -.5cm
\caption{Schematic representation of (a) dijet production with large
rapidity separation $\Delta$ in hadron--hadron collisions, and (b) forward
jet production in deep inelastic scattering.}
\label{fig:feyn}
\end{figure}

Fixed-order QCD perturbation theory fails in  
some asymptotic regimes  where  large logarithms multiply
the coupling constant.  In those regimes resummation of the perturbation 
series  to all orders is necessary to describe many high-energy processes.
The Balitsky-Fadin-Kuraev-Lipatov (BFKL) equation~\cite{bfkl} performs such a
resummation for virtual and real soft gluon emissions in  dijet production at 
large rapidity difference in 
hadron-hadron collisions (see  Figure~\ref{fig:feyn}(a))
and in forward jet
production in lepton-hadron collisions (Figure~\ref{fig:feyn}(b)).  
In the latter case,
resummation leads to the characteristic
BFKL rise 
in the forward jet cross section, $\hat\sigma \sim (x_j/x_{Bj})^\lambda$,
with  $\lambda = 4C_A\ln 2\, \alpha_s/\pi \approx 0.5$.  Similarly, 
in dijet production at hadron colliders 
BFKL resummation gives~\cite{muenav} a
subprocess cross section that increases with rapidity difference as
$\hat\sigma\sim\exp(\lambda \Delta)$,
where $\Delta$ is the rapidity difference of the two jets with comparable
transverse momenta $p_{T1}$ and $p_{T2}$.

Experimental studies of these processes have recently begun at the  
Tevatron $p \bar p$ and HERA $ep$ colliders.  
Tests so far have been inconclusive;  the data tend to lie between
fixed-order QCD and analytic BFKL predictions.  However the 
applicability of these analytic BFKL solutions is limited by the
fact that they implicitly contain integrations over arbitrary numbers
of emitted gluons with arbitrarily large transverse momentum:  there
are no kinematic constraints included.  Furthermore, 
the implicit sum 
over emitted gluons leaves only leading-order kinematics, {\it i.e.}\/
only the momenta of the `external' particles are made explicit.
The absence of kinematic constraints and energy-momentum conservation cannot,
of course, be reproduced in experiments.  While the effects of such constraints
are in principle sub-leading, it is desirable to include them in
predictions to be compared with experimental results.  As we 
will see, kinematic constraints can affect predictions substantially.

%\vspace*{-.5cm}

\section{Monte Carlo Approach to BFKL Physics}

The solution to this problem of lack of kinematic constraints in analytic
BFKL predictions is to unfold the implicit sum over gluons to make the 
gluon sum explicit, and to implement the result in a Monte Carlo
event generator~\cite{os,schmidt}.  This is achieved as follows.
The BFKL equation contains separate integrals over  real and virtual 
emitted gluons.  We can reorganize the equation by combining the 
`unresolved' real emissions --- those with transverse momenta
below some minimum value (in practice chosen to be small
compared to the momentum threshold for measured
jets) --- with the virtual emissions.  Schematically,
we have 
\be
\int_{virtual} + \int_{real} = \int_{virtual+real, unres.} +
\int_{real, res.}
\ee
We  perform
the integration over virtual and unresolved real
emissions  analytically.  The integral containing the 
resolvable real emissions is left explicit.  

We can then solve the  
BFKL equation
by iteration, and we obtain a differential cross section
that contains an explicit sum over emitted gluons along with 
the appropriate phase space factors.  In addition, we obtain
an overall form factor due
to virtual and unresolved emissions. The subprocess cross section is
\be
d\hat\sigma=d\hat\sigma_0\times\sum_{n\ge 0} f_{n}
\ee
where $f_{n}$ is the iterated solution for $n$ real gluons emitted and
contains the overall form factor.
It is then straightforward to implement the result in a Monte Carlo
event generator.   Emitted real (resolved) gluons appear explicitly, 
so that conservation of momentum and energy, as well as
evaluation of parton distributions that multiply $d\hat\sigma$, 
is based on exact kinematics
for each event.  In addition, we include the running of the strong
coupling constant.  See~\cite{os} for further details.

\section{Dijet Production at Hadron Colliders}

At hadron colliders, the BFKL increase in the dijet subprocess cross section 
with rapidity difference is unfortunately washed out by the falling
parton distribution functions (pdfs).  As a result, the BFKL prediction for 
the total cross section is simply a less steep falloff than obtained in 
fixed-order QCD, and tests of this prediction are sensitive to pdf
uncertainties.  A more robust pediction is obtained by noting that 
the emitted gluons (cf. Figure~\ref{fig:feyn}(a)) give rise to
a decorrelation in azimuth between the two leading jets.\cite{many,os}  This 
decorrelation becomes stronger as the rapidity difference $\Delta$ increases
and more gluons are emitted.  In lowest order in QCD, in contrast, the jets
are back-to-back in azimuth and the (subprocess) cross section is 
constant, independent
of $\Delta$.

\begin{figure*}[t]
\psfig{figure=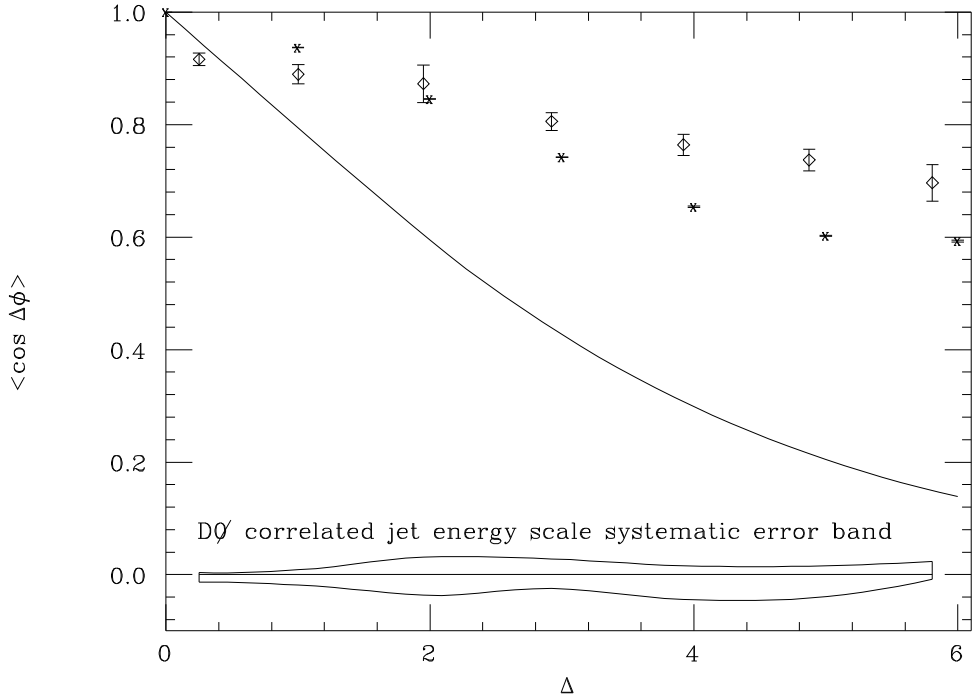,width=16.0cm}
\vskip -.75cm
\caption[]{The azimuthal angle decorrelation in dijet production at the 
Tevatron 
as a function of dijet rapidity difference $\Delta$, for 
jet transverse momentum $p_T>20\ {\rm GeV}$.  
The analytic BFKL solution is shown as a solid curve 
and a preliminary \Dzero\ measurement~\cite{dzeropl} is shown
as diamonds.  Error bars represent statistical and 
uncorrelated systematic errors;  correlated jet energy scale systematics
are shown as an error band.  \label{fig:decor}}
\end{figure*}

This azimuthal decorrelation is illustrated in Figure~\ref{fig:decor}
for dijet production at the Tevatron $p\bar p$ collider~\cite{os}, with 
center of mass energy 1.8 TeV  and jet transverse momentum $p_T>20\ {\rm GeV}$.
The azimuthal angle difference $\Delta\phi$ is defined such that 
that $\cos\Delta\phi=1$ for back-to-back jets.
The solid line shows the analytic BFKl prediction.  The BFKL Monte Carlo
prediction is shown as crosses.  We see that the kinematic constraints
result in a less strong decorrelation due to suppression of 
emitted gluons, and we obtain improved  agreement with
preliminary measurements by the \Dzero\ collaboration~\cite{dzeropl}, 
shown as diamonds in the figure.

\begin{figure*}[t]
%\vskip -1.cm
\psfig{figure=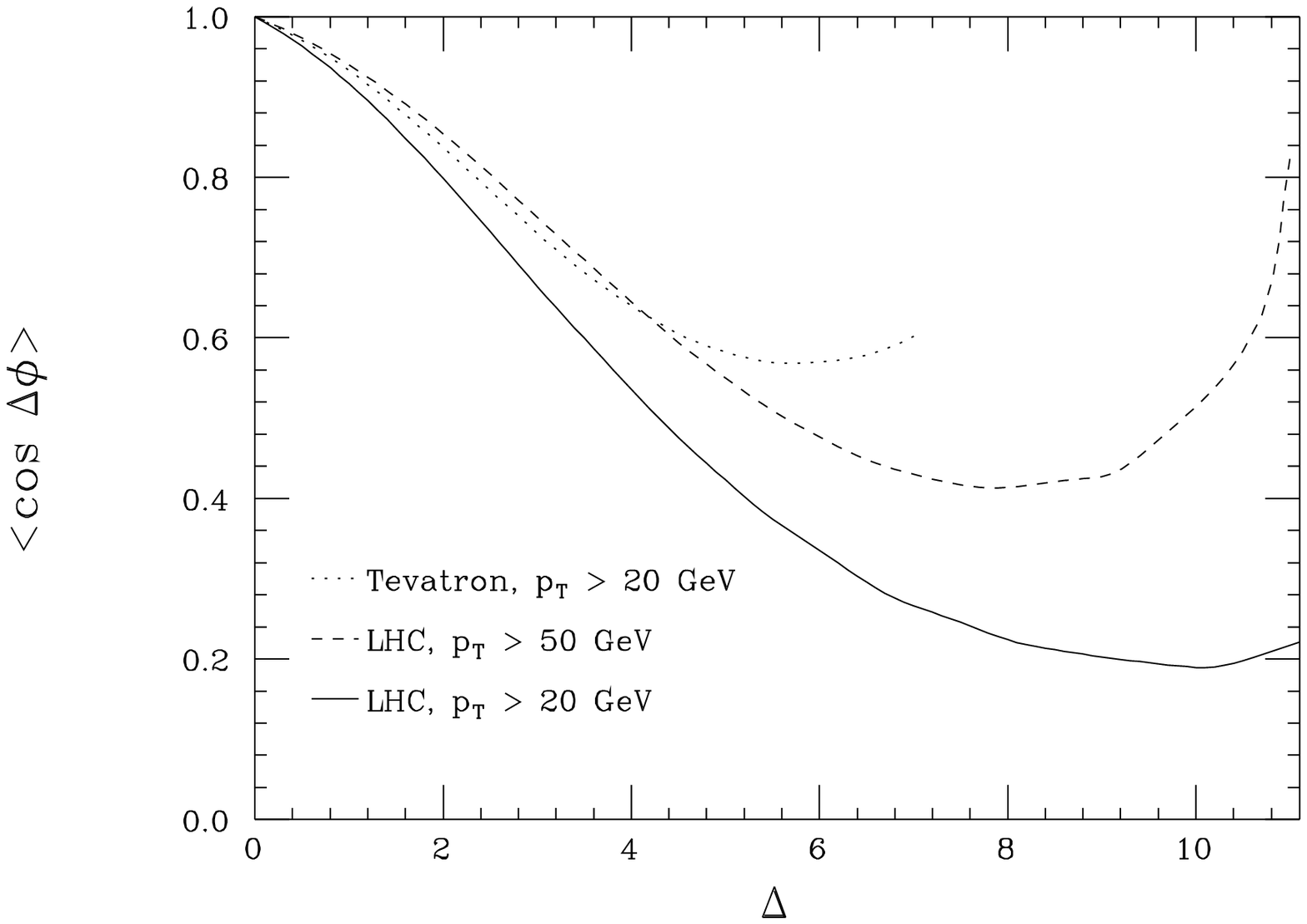,width=16.0cm}
\vskip -.5cm
\caption[]{
The azimuthal angle decorrelation in dijet production at the Tevatron 
($\sqrt{s}=1.8\; \GeV$) and LHC ($\sqrt{s}=14\; \TeV$)
as a function of dijet rapidity difference $\Delta$.  
Dotted curve:  Tevatron, $p_T>20\; \GeV$;
solid curve: LHC, $p_T>20\; \GeV$; dashed curve: LHC, $p_T>50\; \GeV$.   
\label{fig:decorlhc}}
\end{figure*}

The azimuthal decorrelation can also be studied at the LHC $pp$ 
collider~\cite{oslhc}, which has higher rapidity reach than the Tevatron.
Figure~\ref{fig:decorlhc}  compares the decorrelation at the Tevatron 
for $p_T>20\ \GeV$
(dotted curve; same as crosses in Fig.~\ref{fig:decor}) to that at the 
LHC for $p_T>20\ \GeV$ (solid curve) and $p_T>50\ \GeV$ (dashed curve).
We see that at the LHC for $p_T>20\ \GeV$ the decorrelation is stronger 
and reaches to larger rapidities than the Tevatron.  The LHC's
higher center of mass energy ($\sqrt{s}=14\ 
{\rm TeV}$) relative to $p_T$ threshold allows for more emitted gluons, 
and the characteristic BFKL effects are more pronounced.  For the  perhaps
more realistic LHC $p_T$ threshold of $50\ \GeV$,  the 
kinematic suppression is more pronounced, but we still see a
strong decorrelation.
In all three curves we see the suppression of the decorrelation by 
the kinematic 
constraints as $\Delta$ approaches the kinematic limit, where the suppression
of emitted gluons is so strong that the curve turns over and the correlation
begins to return.

In addition to studying the azimuthal decorrelation, one can 
look for the BFKL rise in  dijet cross section with
rapidity difference by considering  ratios of cross sections
at different center of mass energies at fixed $\Delta$.
The idea is to cancel the pdf dependence, leaving the pure
BFKL effect.  This turns out to be rather tricky~\cite{osratio},
because the desired  cancellations occur only at lowest
order, and the kinematic constraints strongly affect the predicted
behavior, not only quantitatively but sometimes 
qualitatively as well~\cite{osratio,oslhc}.

\section{Forward Jet  Production at Lepton--Hadron Colliders}

In deep inelastic scattering at lepton-hadron colliders, the production
of forward jets~\cite{FJ} is subject to the effects of
multiple soft gluon emission just as in dijet production at hadron colliders.
Now the large rapidity separation is between the current and forward
jets; see Fig.~\ref{fig:feyn}(b).
The BFKL equation resums such emissions, and it is 
relatively straightforward 
to adapt the dijet formalism
to calculate the cross section for the production of a forward jet with a given
$k_T$ and longitudinal momentum fraction $x_j \gg \xbj$. In fact there
is a direct correspondence between the variables: $p_{T2} \leftrightarrow 
k_T$ and $\Delta \leftrightarrow \ln(x_j/\xbj)$. In the DIS case the variable
$p_{T1}$ corresponds to the transverse momentum of the $q \bar q$ pair
in the upper `quark box' part of the diagram. In practice this variable
is integrated with the off--shell $\gamma^* g^* \to q \bar q$ amplitude
such that $p_{T1}^2 \sim Q^2$. As a result, it is appropriate to consider
values of $k_T^2$ of the same order, and to consider the (formal) 
kinematic limit $x_j/\xbj \to \infty$, $Q^2$ fixed. In this limit we
obtain the `naive BFKL' prediction 
$\hat \sigma_{\rm jet} \sim (x_j/\xbj)^\lambda$,
the analog of $\hat\sigma_{jj} \sim \exp(\lambda\Delta)$.

\begin{figure*}
\psfig{figure=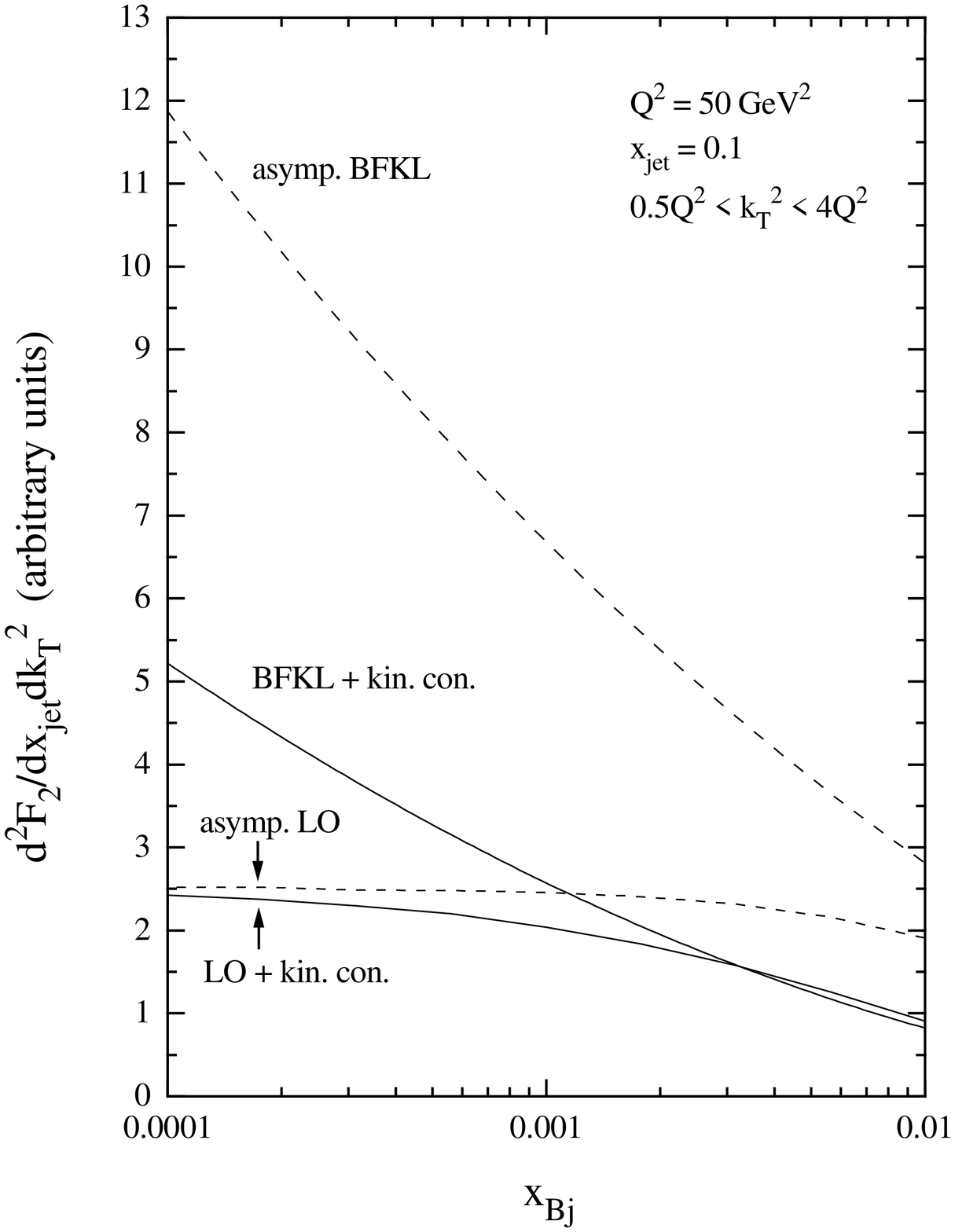,width=16.0cm}
\vskip -1.25cm
\caption{Differential structure function for forward jet production in
$ep$ collisions at HERA. The curves are described in the text.
 \label{fig:djhera}}
\end{figure*}

Figure~\ref{fig:djhera} shows the differential structure function
$\partial^2 F_2/\partial x_j\partial k_T^2$ as a function of $\xbj$ at HERA,
with 
\be
x_j = 0.1, \; \; \; Q^2 = 50\; \GeV^2, \; \; \; Q^2/2 < k_T^2 < 4 Q^2.
\ee
The lower dashed curve is the QCD leading--order prediction from the 
process $\gamma^* \cG \to q \qbar \cG$, with $\cG = g,q$, with no 
overall energy--momentum constraints. This is the analog  of the
$\hat\sigma_{jj} \to\ $constant prediction for dijet production. Note that
here the parton distribution function at the lower end of the ladder
is evaluated at $x = x_j$, independent of $\xbj$. In practice, when 
$\xbj$ is not small we have $x > x_j$ and the cross section is suppressed,
as indicated by the lower solid curve in Fig.~\ref{fig:djhera}. The upper
dashed curve is the asymptotic BFKL prediction with the characteristic
$(x_j/\xbj)^\lambda$ behavior. Finally the upper solid line is the
prediction of the full BFKL Monte Carlo, including kinematic constraints
and pdf dependence. We see a significant suppression of the 
cross section. We emphasise that Fig.~\ref{fig:djhera} corresponds to
`illustrative' cuts and should not be directly compared to the experimental
measurements. Nevertheless, the BFKL--MC predictions do appear to
follow the general trend of the H1 and ZEUS measurements \cite{HERA}.
A more complete study, including realistic experimental cuts and an
assessment of the uncertainty in the theoretical predictions, is under way
and will be reported elsewhere \cite{os3}.

\section{Conclusions}

In summary, we have developed a BFKL Monte Carlo event generator that 
allows us to include the subleading effects such as kinematic constraints
and running of $\alpha_s$.  We have applied this Monte Carlo to 
dijet production at large rapidity separation at the Tevatron and LHC, and
to forward jet production at HERA; the latter work is currently 
being completed.  We found that kinematic constraints, though nominally 
subleading, can be very important.  In particular they lead to suppression
of gluon emission, which in turn suppresses some of the behavior that is 
considered to be characteristic of BFKL physics.  It is clear therefore
that reliable BFKL tests  can only be performed using predictions
that incorporate kinematic constraints.

\section*{Acknowledgements}
Work supported in part by the U.S. Department of Energy,
under grant DE-FG02-91ER40685 and by the U.S. National Science Foundation, 
under grants PHY-9600155 and PHY-9400059.


\section*{References}
\begin{thebibliography}{99}
\bibitem{bfkl}
L.N.~Lipatov, Sov. J. Nucl. Phys. {\bf 23} (1976) 338;
E.A.~Kuraev, L.N.~Lipatov and V.S.~Fadin, 
Sov. Phys. JETP {\bf 45} (1977) 199;
Ya.Ya.~Balitsky and L.N.~Lipatov, Sov. J. Nucl. Phys. {\bf 28} (1978) 822.

\bibitem{muenav}
A.H.~Mueller and H.~Navelet, Nucl. Phys. {\bf B282} (1987) 727.

\bibitem{os}
L.H.~Orr and W.J.~Stirling,
Phys. Rev. {\bf D56} (1997) 5875.

\bibitem{schmidt} C.R.~Schmidt, Phys. Rev. Lett. {\bf 78} (1997) 4531.

\bibitem{many}
V.~Del~Duca and C.R.~Schmidt, Phys. Rev. {\bf D49} (1994) 4510;
W.J.~Stirling, Nucl. Phys. {\bf B423} (1994) 56;
V.~Del~Duca and C.R.~Schmidt, Phys. Rev. {\bf D51} (1995) 215;
V.~Del~Duca and C.R.~Schmidt,
Nucl. Phys. Proc. Suppl. {\bf 39BC} (1995) 137; preprint DESY 94-163 (1994),
presented at the 6th Rencontres de Blois, Blois, France, June 1994.

\bibitem{dzeropl}
\Dzero\ collaboration: S.~Abachi {\it et al.},
Phys. Rev. Lett. {\bf 77} (1996) 595;
\Dzero\ collaboration: presented by Soon Yung Jun
at the  Hadron Collider Physics XII Conference, Stony Brook, June 1997.

\bibitem{oslhc}
L.H.~Orr and W.J.~Stirling,
Phys.~Lett. {\bf B436} (1998) 372.

\bibitem{osratio}
L.H.~Orr and W.J.~Stirling,
Phys.~Lett. {\bf B429} (1998) 135.

\bibitem{FJ}
 A.H.~Mueller, Nucl. Phys. {\bf B}, Proc.
Suppl. {\bf 18 C} (1991) 125;
W.K.~Tang, Phys. Lett. {\bf B278}
(1991) 363;
J.~Bartels {\it et al.}, Z. Phys. {\bf C54} (1992) 635; Phys.
Lett. {\bf B309} (1993) 400; Phys. Lett. {\bf B384} (1996) 300;
J.~Kwieci\'nski, A.D.Martin and P.J. Sutton, Phys. Rev. {\bf D46} (1992) 921; 
 Nucl. Phys. {\bf B}, Proc. Suppl. {\bf 29A} (1992) 67.

\bibitem{HERA} H1 collaboration, these proceedings;
               ZEUS collaboration, these proceedings.

\bibitem{os3}
L.H.~Orr and W.J.~Stirling, in preparation.





\end{thebibliography}
\end{document}